\documentclass[twocolumn,showpacs,preprintnumbers,superscriptaddress,amsmath,amssymb,aps,pre]{revtex4}

\usepackage{graphicx}
\usepackage{dcolumn}
\usepackage{bm}
\usepackage{color}

\begin{document}


\title{Finite-size effect and the components of multifractality in financial volatility}

\author{Wei-Xing Zhou}
 \email{wxzhou@ecust.edu.cn}
 \affiliation{School of Business, East China University of Science and Technology, Shanghai 200237, China} %
 \affiliation{School of Science, East China University of Science and Technology, Shanghai 200237, China} %
 \affiliation{Research Center for Econophysics, East China University of Science and Technology, Shanghai 200237, China} %
 \affiliation{Engineering Research Center of Process Systems Engineering (Ministry of Education), East China University of Science and Technology, Shanghai 200237, China} %
 \affiliation{Research Center on Fictitious Economics \& Data Science, Chinese Academy of Sciences, Beijing 100080, China} %

\date{\today}

\begin{abstract}
Many financial variables are found to exhibit multifractal nature, which is usually attributed to the influence of temporal correlations and fat-tailedness in the probability distribution (PDF). Based on the partition function approach of multifractal analysis, we show that there is a marked finite-size effect in the detection of multifractality, and the effective multifractality is the apparent multifractality after removing the finite-size effect. We find that the effective multifractality can be further decomposed into two components, the PDF component and the nonlinearity
component. Referring to the normal distribution, we can determine the PDF component by comparing the effective multifractality of the original time series and the surrogate data that have a normal distribution and keep the same linear and nonlinear correlations as the original data. We demonstrate our method by taking the daily volatility data of Dow Jones Industrial Average from 26 May 1896 to 27 April 2007 as an example. Extensive numerical experiments show that a time series exhibits effective multifractality only if it possesses nonlinearity and the PDF has impact on the effective multifractality only when the time series possesses nonlinearity. Our method can also be applied to judge the presence of multifractality and determine its components of multifractal time series in other complex systems.
\end{abstract}


\maketitle


\section{Introduction}
\label{S1:Intro}

Turbulent flows exhibit multifractal nature. Two main pictures of multifractal analysis in the turbulence literature deal with the velocity fluctuations using the structure function approach and the energy dissipation rates using the partition function approach \cite{Frisch-1996}. The difference between the two pictures is remarkable and there are a lot of efforts trying to relate the two sequences of scaling exponents. In addition, the power-law dependence of the structure function is more or less not perfect.

It is well known that there are many similarities between turbulent flows (velocity time series) and financial markets (equity prices) \cite{Ghashghaie-Breymann-Peinke-Talkner-Dodge-1996-Nature,Mantegna-Stanley-1996-Nature,Mantegna-Stanley-1997-PA,Mantegna-Stanley-2000}. For instance, multifractality is one of the most elusive stylized facts in financial markets \cite{Ghashghaie-Breymann-Peinke-Talkner-Dodge-1996-Nature,Mantegna-Stanley-1996-Nature,Mandelbrot-1999-SA}. Many different methods have been applied to characterize the hidden multifractal behavior of different financial variables, such as the fluctuation scaling analysis \cite{Eisler-Kertesz-Yook-Barabasi-2005-EPL,Eisler-Kertesz-2007-EPL,Jiang-Guo-Zhou-2007-EPJB}, the structure function method \cite{Ghashghaie-Breymann-Peinke-Talkner-Dodge-1996-Nature,Vandewalle-Ausloos-1998-EPJB,Ivanova-Ausloos-1999-EPJB,Schmitt-Schertzer-Lovejoy-1999-ASMDA,Schmitt-Schertzer-Lovejoy-2000-IJTAF,Calvet-Fisher-2002-RES,Ausloos-Ivanova-2002-CPC,Gorski-Drozdz-Speth-2002-PA,AlvarezRamirez-Cisneros-IbarraValdez-Soriano-2002-PA,Balcilar-2003-EMFT,Lee-Lee-2005a-JKPS,Lee-Lee-Rikvold-2006-PA}, the multifractal detrended fluctuation analysis (MF-DFA) \cite{Kantelhardt-Zschiegner-KoscielnyBunde-Havlin-Bunde-Stanley-2002-PA,Matia-Ashkenazy-Stanley-2003-EPL,Kwapien-Oswiecimka-Drozdz-2005-PA,Lee-Lee-2005b-JKPS,Oswiecimka-Kwapien-Drozdz-2005-PA,Moyano-Souza-Queiros-2006-PA,Jiang-Ma-Cai-2007-PA,Lee-Lee-2007-PA,Lim-Kim-Lee-Kim-Lee-2007-PA,Su-Wang-Huang-2009-JKPS}, the partition function method \cite{Sun-Chen-Wu-Yuan-2001-PA,Sun-Chen-Yuan-Wu-2001-PA,Ho-Lee-Wang-Chuang-2004-PA,Wei-Huang-2005-PA,Gu-Chen-Zhou-2007-EPJB,Du-Ning-2008-PA,Zhuang-Yuan-2008-PA,Wei-Wang-2008-PA,Zhou-2010-cnJMSC,Jiang-Zhou-2008a-PA,Su-Wang-2009-JKPS}, the multiplier method \cite{Jiang-Zhou-2007-PA}, and the wavelet transform approaches \cite{Struzik-Siebes-2002-PA,Turiel-Perez-Vicente-2003-PA,Turiel-Perez-Vicente-2005-PA,Oswiecimka-Kwapien-Drozdz-Rak-2005-APPB}, some of which are borrowed from the multifractal analysis of turbulence data. When the return time series is concerned, the structure function method and the MF-DFA method apply. For volatility time series, the partition function approach and the multiplier approach can be utilized.

The extracted multifractal nature has potential applications in financial engineering. Some researchers report that the width of the estimated multifractal spectrum is correlated to the price fluctuation in the future and thus can be used to predict price fluctuations \cite{Sun-Chen-Yuan-Wu-2001-PA,Wei-Huang-2005-PA,Su-Wang-2009-JKPS}. In a similar vein, the so-called {\em{multifractal volatility}} has been introduced to measure stock market risks, which can be used to estimate the dynamic Value-at-Risk of an asset \cite{Wei-Wang-2008-PA}. There is also empirical evidence suggesting that multifractal analysis can be used to quantify the degree of inefficiency of markets in the sense that more developed stock markets have weaker multifractality \cite{Zunino-Tabak-Figliola-Perez-Garavaglia-Rosso-2008-PA,Zunino-Figliola-Tabak-Perez-Garavaglia-Rosso-2009-CSF} and an emerging market evolves to be more efficient with narrowing singularity width \cite{Wang-Liu-Gu-2009-IRFA}.

However, understanding the origin of multifractality in financial markets and its components is still a subtle and open problem. For instance, standard multifractal analysis of an exactly monofractal financial model gives artificial multifractal behaviors \cite{Bouchaud-Potters-Meyer-2000-EPJB}. It is usually argued that the fat tails and the long-term power-law correlations are two possible sources of multifractal nature in financial time series \cite{Kantelhardt-Zschiegner-KoscielnyBunde-Havlin-Bunde-Stanley-2002-PA}. It is obvious that possessing only linear correlations is not sufficient for the presence of multifractality and a nonlinear process with long-memory is necessary to have multifractality \cite{Saichev-Sornette-2006-PRE}. Numerical investigations show that the reported multifractal nature in many real cases is stemmed from the large fluctuations of prices \cite{Lux-2004-IJMPC}.

Empirically, the problem has been studied based on the multifractal detrended analysis of financial returns \cite{Zhou-2009-EPL}. One conventional measure for quantifying the degree of multifractality is the width of singularity spectrum
\begin{equation}
 \Delta\alpha = \alpha_{\max} - \alpha_{\min}.
 \label{Eq:dA:def}
\end{equation}
Alternative measures are also used in some cases \cite{Zunino-Tabak-Figliola-Perez-Garavaglia-Rosso-2008-PA,deSouza-Queiros-2009-CSF}. In this way, one can quantitatively determine the contribution of the correlation and PDF components \cite{Jin-Lu-2006-INCB,deSouza-Queiros-2009-CSF}. In order to investigate the influence of temporal correlations in return series, one can randomly shuffle the original data and compare their singularity spectra \cite{Matia-Ashkenazy-Stanley-2003-EPL,Oswiecimka-Kwapien-Drozdz-Rak-2005-APPB,Lee-Lee-2005b-JKPS,Kwapien-Oswiecimka-Drozdz-2005-PA,Jin-Lu-2006-INCB,Kumar-Deo-2009-PA,deSouza-Queiros-2009-CSF}. All these studies show that the shuffled data have non-shrinking singularity width $\Delta\alpha_{\rm{SF}}$, comparable to the original width $\Delta\alpha$. These observations imply that the heavy-tailed distribution of the returns has a crucial impact on the singularity width. To understand the impact of the distribution, one can either remove large returns \cite{Oh-Eom-Havlin-Jung-Wang-Stanley-Kim-2010-PRE} or generate surrogate data having a Gaussian distribution while keeping the linear correlation of the original data \cite{Norouzzadeh-Rahmani-2006-PA,Lim-Kim-Lee-Kim-Lee-2007-PA,Su-Wang-Huang-2009-JKPS}. In addition, the temporal correlations can be further separated into linear and nonlinear components, which have different effects on the multifractal spectrum \cite{Zhou-2009-EPL}.

In this work, based on the partition function method of multifractal analysis, we propose that the apparent multifractality can be decomposed into three components caused by the nonlinear correlation, the linear correlation (long-term power-law memory) and the fat-tailed PDF, which can be characterized by the following expression:
\begin{equation}
 \Delta\alpha =
 \Delta\alpha_{\rm{NL}}+\Delta\alpha_{\rm{LM}}+\Delta\alpha_{\rm{PDF}}.
 \label{Eq:Decomposition}
\end{equation}
We find for the first time that the linear correlation component $\Delta\alpha_{\rm{LM}}$ is the outcome of the finite-size effect and thus the intrinsic multifractal nature is characterized by the effective multifractality $\Delta\alpha_{\rm{eff}}$ composing of the nonlinearity component $\Delta\alpha_{\rm{NL}}$ and the PDF component $\Delta\alpha_{\rm{PDF}}$. These properties are very different from those of the MF-DFA results in Ref.~\cite{Zhou-2009-EPL}.

We will develop a systemic procedure to quantitatively determine the components of multifractality. The analysis is carried out based on the daily volatility data of the Dow Jones Industrial Average (DJIA) from 26 May 1896 to 27 April 2007 (totally 30147 trading days). The logarithmic returns is defined as
\begin{equation}
 r(t) = \ln P_t-\ln P_{t-1},
 \label{Eq:rt}
\end{equation}
where $P_t$ is the price at time $t$, and the volatility is computed as the absolute value of $r(t)$:
\begin{equation}
 v(t)=|r(t)|.
 \label{Eq:vt}
\end{equation}
This simple definition is widely adopted in the Econophysics community.

This paper is organized as follows. Section \ref{S1:PF:MF} gives a brief description of the partition function approach for multifractal analysis and its numerical implementation. Section \ref{S1:FSE} designs a numerical procedure to confirm the presence of finite-size effect in the detection of multifractality in time series with a prescribed probability distribution. In Sec.~\ref{S1:3components}, the three multifractal components are determined for the daily DJIA volatility, where the nonlinear correlation component and the fat-tail component are determined by comparing with surrogate time series with normal distributions. The impact of the fat-tailedness of the probability distribution on the apparent multifractality is further investigated in Sec.~\ref{S1:ImpactPDF} using Student's $t$ distributions and the Weibull distributions. We find that the fat-tailed PDF per se has no effect on the multifractality and its impact is due to its coupling with the nonlinear correlations. We finally provide some concluding
remarks in Sec.~\ref{S1:Summary}.

\section{Multifractal analysis based on partition functions}
\label{S1:PF:MF}

The multifractal analysis in this paper is based on the partition function formalism \cite{Halsey-Jensen-Kadanoff-Procaccia-Shraiman-1986-PRA}. Our procedure is slightly different from the conventional algorithm in the calculation of the partition functions. Consider a time series $\{v_i: i=1,2,\cdots,N\}$. For a given integer time scale $\ell$, we choose randomly $m$ integers $\{j_i: i=1,2,\cdots,m\}$ uniformly distributed in $[1,N-\ell+1]$. Then the measure enclosed in the interval $[j_i,j_i+\ell-1]$ is calculated
\begin{equation}
  E_i(\ell) =  \left. \sum_{k=0}^{\ell-1} v_{j_i+k}\right/ \sum_{k=1}^N v_k,
  \label{Eq:MF:Ei}
\end{equation}
The $q$-th partition function is calculated as follows
\begin{equation}
  M_q(\ell) = \frac{N}{m\ell} \sum_{i=1}^m [E_i(\ell)]^q.
  \label{Eq:MF:Mq}
\end{equation}
If the time series is power-law long-term correlated, we have
\begin{equation}
  M_q(\ell) \sim \left(\ell/N\right)^{\tau(q)}.
  \label{Eq:MF:tau}
\end{equation}
If the scaling exponent $\tau(q)$ is a nonlinear function of $q$, the time series possess multifractal nature. The singularity strength and its spectrum can be computed according to the Legendre transform of $\tau(q)$ \cite{Halsey-Jensen-Kadanoff-Procaccia-Shraiman-1986-PRA},
\begin{equation}
  \alpha(q) = \tau'(q) \ {\mathrm{and}} \
  f(\alpha) = q\alpha-\tau.
  \label{Eq:MF:alpha:f}
\end{equation}
The advantage of using Eq.~(\ref{Eq:MF:Mq}) is twofold. First, it gives better statistics when $\ell$ is close to $N$. Second, it does not need to deal with the situation in which $N/\ell$ is not integer.

\section{Finite-size effect}
\label{S1:FSE}

The finite-size effect in the detection of multifractality has been documented for several uncorrelated time series. One example is time series having an exponential distribution, whose $q$-order moment can be derived as \cite{vonHardenberg-Thieberger-Provenzale-2000-PLA}
\begin{equation}
 M_q(\ell) = \Gamma(\ell+q)L/\Gamma(\ell+1).
 \label{Eq:Mq:Exp}
\end{equation}
When the length $L$ of the time series is small, a spurious multifractal spectrum is detected, where the singularity width $\Delta\alpha$ is significantly larger than zero. Only for sufficiently long time series with $L\gg0$ and for $\ell\gg q$, the moment is approximated by $M_q(\ell)\sim \ell^{q-1}$. It follows immediately that $\tau=q-1$, $\alpha=1$ and $f(\alpha)=1$. Therefore, the finite-size effect results in a spurious detection of multifractality from a monofractal signal if the length of the signal is short or the ``scaling range'' locates at small scales.

Another example is exponentially truncated L\'evy flights with characteristic parameter $\gamma$, which exhibit a bifractal behavior such that $\tau(q)=q/\gamma-1$ when $0<q<\gamma$ and $\tau(q)=0$ when $q\geqslant\gamma$ \cite{Nakao-2000-PLA}. The multifractal spectrum $f(\alpha)$ shrinks into two points $(0,0)$ and $(1/\gamma,1)$. Extensive numerical experiments using uncorrelated time series obeying $q$-Gaussian distributions with different tail exponents unveil a convergence to monofractalilty in the Gaussian attraction basin and to bifractality in the L\'evy attraction basin \cite{Drozdz-Kwapien-Oswiecimka-Speth-2009-XXX}, which is consistent with the analytic results for exponentially truncated L\'evy flights \cite{Nakao-2000-PLA}.

More generally, according to the law of large numbers, the sum of $\ell$ i.i.d. random variables with mean $\mu$ and standard deviation $\sigma$ converges to $\ell\mu$ with much smaller fluctuations \cite{Hsu-Robbins-1947-PNAS}. Let $v_i>0$ be a sequence of independent random variables with mean $\mu$ and variance $\sigma^2$ and
\begin{equation}
 V=\sum_{i=1}^{\ell} v_i.
 \label{Eq:SI:V}
\end{equation}
Denoting that
\begin{equation}
 x_i= v_i-\mu
 \label{Eq:SI:xi}
\end{equation}
and
\begin{equation}
 X = \sum_{i=1}^{\ell} x_i,
 \label{Eq:SI:X}
\end{equation}
we have
\begin{equation}
 V = X + \ell\mu.
 \label{Eq:SI:V:X}
\end{equation}
According to the law of large numbers \cite{Hsu-Robbins-1947-PNAS},
$X$ converges to $0$, if
\begin{equation}
 \int x dF(x) =0 \ \ {\rm{and}} \ \  \int x^2 dF(x) <\infty,
 \label{Eq:SI:xdF}
\end{equation}
where $F(x)$ is the cumulative distribution of $x$. Hence, $V$
converges to $\ell\mu$. In addition, according to the Central Limit
Theorem, the distribution of $X$ is Gaussian:
\begin{equation}
 P(X) = \frac{1}{\sqrt{2\pi\ell}\sigma}\exp\left[-\frac{(X-\ell\mu)^2}{2\ell\sigma^2}\right].
 \label{Eq:SI:Pr:X}
\end{equation}
We have
\begin{eqnarray}
 M_q(\ell) &=& \frac{N}{\ell} \int V^q P(V) dV  \nonumber\\
           &=& \frac{N}{\ell} \int (X+\ell\mu)^q P(X) dX  \nonumber\\
           &\approx& N \ell^{q-1} \int P(X) dX  \nonumber\\
           &=& N \ell^{q-1}
 \label{Eq:SI:Mq}
\end{eqnarray}
where $P(X)dX=P(V)dV$ and $X\ll\ell\mu$ for most $X$ are used. It follows immediately that $\tau=q-1$, $\alpha=1$ and $f(\alpha)=1$. It indicates that the time series is monofractal and its ``singularity spectrum'' shrinks to a single point $(\alpha,f)=(1,1)$. A direct consequence is that any degree of multifractality observed in real uncorrelated time series is caused by a finite-size effect.

However, the impact of linear long-term correlations on the finite-size effect has not been investigated. It is well-known that financial volatility exhibits strong long-term memory \cite{Ding-Granger-1996-JEm,Mantegna-Stanley-2000}. The Hurst index of the daily DJIA volatility is $H=0.80$. In order to show that there is a finite-size effect in the detection of multifractality introduced by linear correlations in the volatility time series, we generate surrogate data that have the same probability distribution as the original time series of volatility, which can be done with the transformation method \cite{Press-Teukolsky-Vetterling-Flannery-1996}. We first construct the empirical cumulative distribution $F(v)$ of volatility, which is the occurrence frequency of volatility $v(t)$ less than $v$. A sequence of random numbers $\{x_i: i=1,2,\cdots,L\}$ are drawn from a uniform distribution. In order not to introduce very large values in the surrogate data of volatility caused by the edge effect, we perform the following map
\begin{equation}
 y_i = \frac{(x_i-\min\{x_j\})(1-\min\{F\})} {\max\{x_j\}-\min\{x_j\}} + \min\{F\}.
 \label{Eq:xi}
\end{equation}
The sequence of numbers $\{z_i: i=1,2,\cdots,L\}$ are calculated according to the following transformation
\begin{equation}
 z_i = F^{-1}(y_i),
 \label{Eq:zi}
\end{equation}
which can be done numerically by linear or spline interpolations. The sample $\{z_i: i=1,2,\cdots,L\}$ has the same distribution as the volatility $\{v(t): t=1,2,\cdots,N\}$.

We then introduce long memory (linear correlations) in the time series $\{z_i\}$ using an improved amplitude adjusted Fourier transform (IAAFT) algorithm\cite{Schreiber-Schmitz-1996-PRL}, which is based on a simple iteration scheme and an improved version of the amplitude adjusted Fourier transform algorithm \cite{Theiler-Eubank-Longtin-Galdrikian-Farmer-1992-PD}. The volatility data $\{v(t):t=1,2,\cdots,N\}$ are sorted resulting in a new sequence $\{s_N\}$, and we obtain the squared amplitudes of the Fourier transform of $\{s_N\}$, denoted as $\{S_k^2\}$. The initial sequence $\{s_N^{(0)}\}$ of the iteration is a random shuffle of $\{s_N\}$. In the $i$-th iteration, the squared amplitudes $\{S_k^{2,(i)}\}$ of the Fourier transform of $\{s_N^{(0)}\}$ are obtained and replaced by $\{S_k^2\}$, which are transformed back, and then the the resulting series are replaced by $\{s_N\}$ but keeping the rank order.

\begin{figure}[htb]
\centering
\includegraphics[width=8cm]{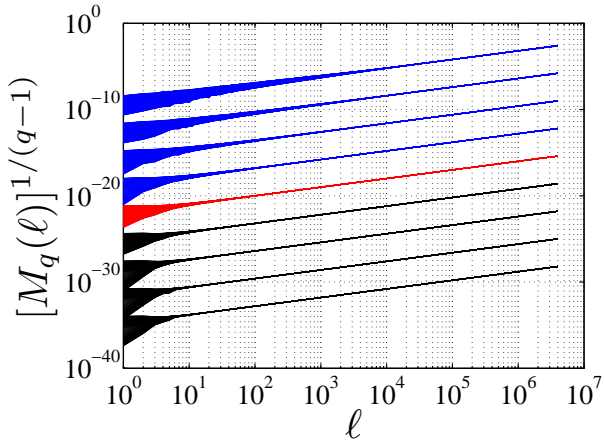}
\caption{\label{Fig:MFComp:Volatility:FiniteSize:M} (Color online) Dependence of $M_q(\ell)^{1/(q-1)}$ as a function of $\ell$ for different surrogate time series with different linear correlations. The Hurst index ranges from 0.9 to 0.1 from the top to the bottom.}
\end{figure}

In our numerical experiments, we have investigated different Hurst indexes with $H=0.1,0.2,\cdots,0.9$. The length $L$ of the surrogate data ranges from $10^3$ to $10^7$. Figure \ref{Fig:MFComp:Volatility:FiniteSize:M} shows the dependence of $M_q(\ell)^{1/(q-1)}$ as a function of $\ell$ for different surrogate time series with different linear correlations. According to Eq.~(\ref{Eq:MF:tau}), the slope of linear lines in Fig.~\ref{Fig:MFComp:Volatility:FiniteSize:M} is $\tau(q)/(q-1)$. For monofractal time series, these lines should be parallel and have identical slopes. For short time series, the scaling range locates on the left of plot and the lines are not parallel. Only when the time series is sufficiently long, the monofractality can be detected. More data points are needed to reach a decisive conclusion for strongly correlated time series with large Hurst index $H$.

\begin{figure}[htb]
\centering
\includegraphics[width=8cm]{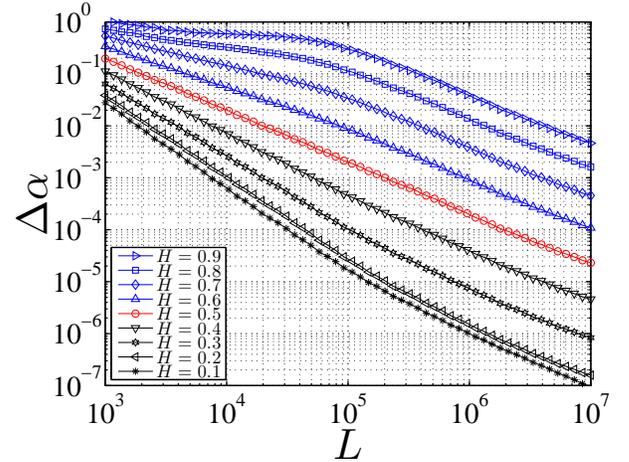}
\caption{\label{Fig:MFComp:Volatility:FiniteSize:Dalpha} (Color online) Quantification of the finite-size effect in the detection of multifractality.}
\end{figure}

For each surrogate time series with Hurst index $H$ and size $L$, a multifractal spectrum is determined within the scaling range $[L/60, L/3]$. Figure~\ref{Fig:MFComp:Volatility:FiniteSize:Dalpha} illustrates the finite-size effect in the detection of multifractality. For a given $H$, the singularity width $\Delta\alpha$ is a decreasing function of the series length $L$. It is shown that the observed non-vanishing singularity spectrum in the surrogate data with only linear correlations is just the outcome of a finite-size effect,
\begin{equation}
 \Delta\alpha_{\rm{LM}}=\Delta\alpha_{\rm{FSE}}.
 \label{Eq:dA:LM:FSE}
\end{equation}
Therefore we can define an effective singularity width
\begin{equation}
 \Delta\alpha_{\rm{eff}} = \Delta\alpha-\Delta\alpha_{\rm{LM}}=\Delta\alpha_{\rm{NL}}+\Delta\alpha_{\rm{PDF}}
 \label{Eq:dA:eff}
\end{equation}
which accounts for the contribution of the nonlinear correlations and the fat tails to the apparent multifractal spectrum width.

\begin{figure}[htb]
 \centering
 \includegraphics[width=8cm]{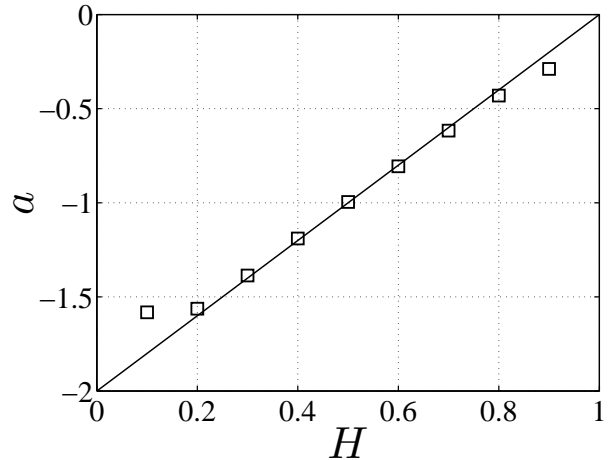}
 \caption{\label{Fig:SI:MFComp:Volatility:FiniteSize:a:H} Dependence of the power-law exponent $a$ on the Hurst index $H$.}
\end{figure}

According to Fig.~\ref{Fig:MFComp:Volatility:FiniteSize:Dalpha}, there is a power-law dependence between the singularity width $\Delta\alpha(H,L)$ and the length $L$ for each Hurst index $H$ when $L$ is not large:
\begin{equation}
 \Delta\alpha(H,L) \sim L^{a}
 \label{Eq:SI:FSE:dA1}
\end{equation}
where $a$ is a function of $H$. For each $H$, we calculate $a$ in the scaling range $[1000,138950]$. The dependence of the exponent $a$ on the Hurst index $H$ is shown in Fig.~\ref{Fig:SI:MFComp:Volatility:FiniteSize:a:H}. We find a nice linear relation
\begin{equation}
 a = 2H-2
 \label{Eq:SI:FSE:a:H}
\end{equation}
except for $H=0.1$. It is interesting to note that, for $H=0.5$, we have $a=-1$ and thus $\Delta\alpha(H,L)\propto L^{-1}$.

\begin{figure}[htb]
 \centering
 \includegraphics[width=8cm]{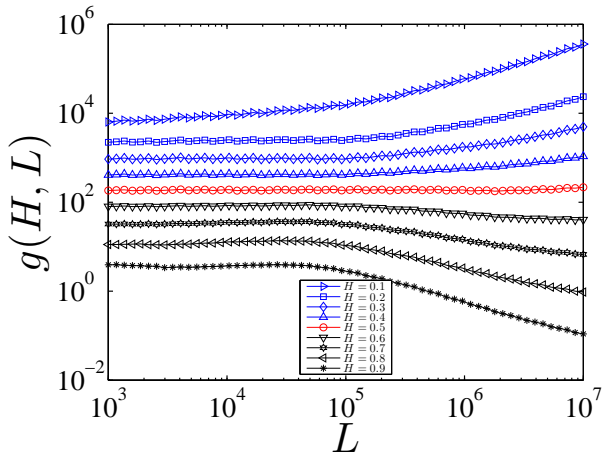}
 \caption{\label{Fig:SI:MFComp:Volatility:FiniteSize:g} (Color online) Dependence of $g(H,L)$ as a function of $L$ for different Hurst index $H$.}
\end{figure}

We assume that $\Delta\alpha(H,L)$ can be factorized as follows
\begin{equation}
 \Delta\alpha(H,L) = g(H,L) L^{a}.
 \label{Eq:SI:FSE:dA2}
\end{equation}
Figure~\ref{Fig:SI:MFComp:Volatility:FiniteSize:g} plots $g(H,L)$ as a function of $L$ for each $H$. We find that $g(H,L)$ is almost independent of $L$, in the scaling range. For the special case with the Hurst index $H=0.5$, it is observed that the $g(H,L)$ curve is horizontal in the whole range of $L$.

\begin{figure}[htb]
 \centering
 \includegraphics[width=8cm]{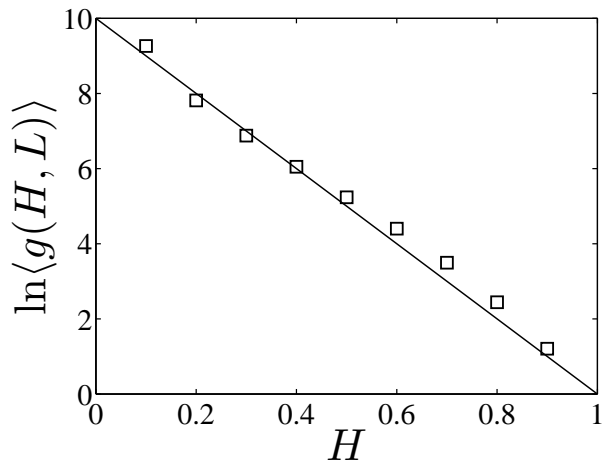}
 \caption{\label{Fig:SI:MFComp:Volatility:FiniteSize:g:ave} Exponential dependence of $\langle g(H,L) \rangle$ with respect to the Hurst index $H$.}
\end{figure}

According to Fig.~\ref{Fig:SI:MFComp:Volatility:FiniteSize:g}, we can assume that $g(H,L)$ is independent of $L$ in the scaling range. We calculate the average $\langle g(H,L) \rangle$ of $g(H,L)$ in the scaling range
\begin{equation}
 \langle g(H,L) \rangle = \frac{1}{n}\sum_{L\in[1000,138950]} g(H,L)
 \label{Eq:SI:FSE:ave:g}
\end{equation}
where $n$ is the number of scales $L$ in the scaling range. Figure~\ref{Fig:SI:MFComp:Volatility:FiniteSize:g:ave} plots $\ln\langle g(H,L) \rangle$ as a function of $H$. We observe a linear relation
\begin{equation}
 \ln \langle g(H,L) \rangle = 10-10H,
 \label{Eq:SI:FSE:ave:g:H}
\end{equation}
which is obtained based on the linear least-squares regression.

Combining Eqs.~(\ref{Eq:SI:FSE:a:H}), (\ref{Eq:SI:FSE:dA2}) and
(\ref{Eq:SI:FSE:ave:g:H}), we obtain
\begin{equation}
 \Delta\alpha(H,L) \approx L^{-2(1-H)} e^{10(1-H)}
 \label{Eq:FSE}
\end{equation}
when $L$ is not large. It is interesting to note that $\Delta\alpha(H,L)\propto L^{-1}$ for $H=0.5$ in the whole range of $L$. Note that this expression (\ref{Eq:FSE}) is valid only for not large $L$. Also, in the quantitative assessment of finite-size effect in correlated signals, one should use Fig.~\ref{Fig:MFComp:Volatility:FiniteSize:Dalpha}, rather than the above expression. For other applications, one need to conduct numerical simulations to produce a finite-size effect figure as Fig.~\ref{Fig:MFComp:Volatility:FiniteSize:Dalpha} based on the sample distribution under investigation.

\section{Determining the singularity width components}
\label{S1:3components}

\begin{figure}[htb]
\centering
\includegraphics[width=8cm]{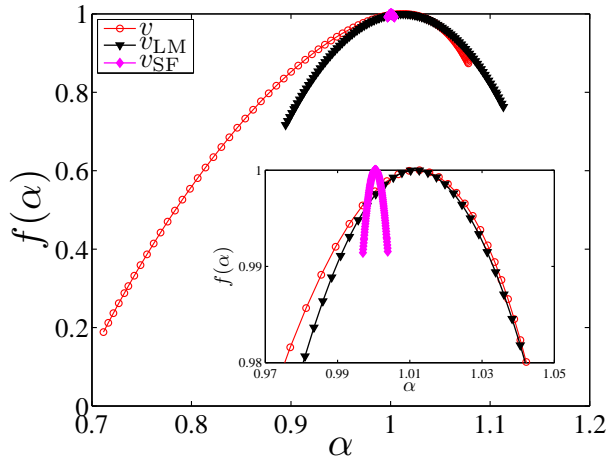}
\caption{\label{Fig:MFComp:Volatility:Correlation} (Color online) Determination of the apparent singularity width $\Delta\alpha$, the finite-size effect component $\Delta\alpha_{\rm{FSE}}$, and the effective singularity width $\Delta\alpha_{\rm{eff}}$ of the daily DJIA volatility. The inset is the amplification of the main plot around $(\alpha,f)=(1,1)$.}
\end{figure}

We now try to determine the components of the apparent singularity width $\Delta\alpha$. Figure~\ref{Fig:MFComp:Volatility:Correlation} shows the multifractal spectrum of the original daily DJIA volatility time series. The apparent singularity width is found to be $\Delta\alpha=0.367$.

In order to determine the finite-size effect component caused by the linear long-term correlations in the volatility series, we use the iteration scheme \cite{Schreiber-Schmitz-1996-PRL} to generate surrogate data $v_{\rm{LM}}$ which have the same distribution and linear long-term correlations as the original volatility time series, while any underlying nonlinear correlations have been eliminated. One hundred surrogate time series $v_{\rm{LM}}$ have been generated and the average multifractal spectrum is shown in Fig.~\ref{Fig:MFComp:Volatility:Correlation}. The finite-size effect component is found to be
\begin{equation}
 \Delta\alpha_{\rm{FSE}}=\Delta\alpha_{\rm{LM}} = 0.219\pm 0.038.
 \label{Eq:Original:dA:FSE}
\end{equation}
For the shuffled data $v_{\rm{SF}}$, the resulting multifractal spectrum shrinks to a point $(\alpha,f)=(1,1)$ as expected and the singularity spectrum width $\Delta\alpha_{\rm{SF}} = 0.007\pm0.001$ is close to zero, which confirms that the fat-tailedness of PDF alone cannot produce any multifractality. This finding is very different from that in the multifractal detrended fluctuation analysis of financial returns. It follows that the effective width of singularity is
\begin{equation}
 \Delta\alpha_{\rm{eff}}=\Delta\alpha-\Delta\alpha_{\rm{FSE}}=0.148\pm 0.038,
 \label{Eq:Original:dA:eff}
\end{equation}
which is significantly smaller than the apparent singularity width $\Delta\alpha$.

In order to determine the PDF component $\Delta\alpha_{\rm{PDF}}$, we need to generate surrogate data in which the linear and nonlinear correlations are preserved while the original PDF is replaced by a reference PDF such that the PDF of the surrogate data has no impact on the resulting multifractality. It is not irrational to assume that the PDF component of Gaussian signals is negligible, that is, $\Delta\alpha_{\rm{norm,PDF}}=0$. We generate surrogate data $r_{\rm{norm}}$ for the returns $r(t)$ which are drawn from a normal distribution
\begin{equation}
 p(r_{\rm{norm}}) = \frac{1}{\sqrt{2\pi}\sigma}e^{-\frac{(r_{\rm{norm}}-\mu)^2}{2\sigma^2}},
 \label{Eq:pdf:Gaussian}
\end{equation}
where $\mu=0.00021$ and $\sigma=0.011$ (in units of one trading day) are the sample mean and standard deviation of the daily DJIA returns $r(t)$. The surrogate volatility is thus $v_{\rm{norm,SF}}=|r_{\rm{norm}}|$, which has no temporal correlation and can be regarded as being shuffled.

The random numbers $v_{\rm{norm,SF}}$ are rearranged to have the same rank ordering as $v(t)$ so that the surrogate time series $v_{\rm{norm}}(t)$ preserves the linear and nonlinear correlations of the original volatility $v(t)$ \cite{Bogachev-Eichner-Bunde-2007-PRL,Zhou-2008-PRE,Zhou-2009-EPL}. The approach is to replace the raw data by random numbers drawn from a prescribed distribution, which is described as follows. For a given distribution, we generate a sequence of random numbers $\{x_0(t):t=1,2,\cdots,N\}$, which are rearranged such that the resulting series $\{x(t):t=1,2,\cdots,N\}$ has the same rank ordering as the volatility series $\{v(t):t=1,2,\cdots,N\}$. In other words, $x(t)$ should rank $n$ in the sequence $\{x(t):t=1,2,\cdots,N\}$ if and only if $v(t)$ ranks $n$ in the $\{v(t):t=1,2,\cdots,N\}$ sequence \cite{Bogachev-Eichner-Bunde-2007-PRL,Zhou-2008-PRE,Zhou-2009-EPL}.

\begin{figure}[htb]
 \centering
 \includegraphics[width=8cm]{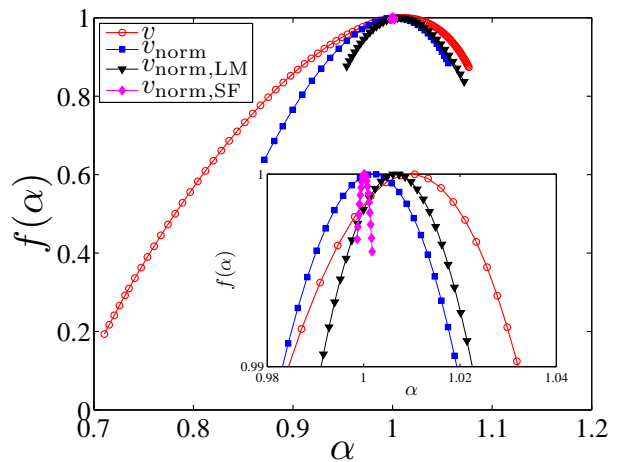}
 \caption{\label{Fig:MFComp:Volatility:PDF:norm} (Color online) Determination of the PDF component $\Delta\alpha_{\rm{PDF}}$ and the nonlinearity component $\Delta\alpha_{\rm{NL}}$ of the original volatility with reference to Gaussian surrogates. The inset is the amplification of the main plot around $(\alpha,f)=(1,1)$.}
\end{figure}

We generate 100 surrogates $v_{\rm{norm}}(t)$. For each time series $v_{\rm{norm}}(t)$, we further destroy its nonlinear correlations but keep its linear correlations to generate $v_{\rm{norm,LM}}(t)$ based on the iterated amplitude adjusted Fourier transformation algorithm \cite{Schreiber-Schmitz-1996-PRL}. The average multifractal spectra for $v_{\rm{norm}}(t)$, $v_{\rm{norm,LM}}(t)$, and $v_{\rm{norm,SF}}(t)$ are illustrated in Fig.~\ref{Fig:MFComp:Volatility:PDF:norm}. We find that, in the Gaussian surrogate case, the apparent singularity width is
\begin{equation}
 \Delta\alpha_{\rm{norm}}=0.185\pm0.003
\end{equation}
for $v_{\rm{norm}}(t)$, its finite-size effect component is
\begin{equation}
 \Delta\alpha_{\rm{norm,FSE}}=\Delta\alpha_{\rm{norm,LM}}=0.118\pm0.028
\end{equation}
for $v_{\rm{norm,LM}}(t)$, and
\begin{equation}
 \Delta\alpha_{\rm{norm,SF}}=0.003\pm0.001
\end{equation}
for $v_{\rm{norm,SF}}(t)$. Therefore, the effective width of singularity
for the Gaussian surrogates is
\begin{equation}
 \Delta\alpha_{\rm{norm,eff}}=\Delta\alpha_{\rm{norm}}-\Delta\alpha_{\rm{norm,FSE}}=0.067\pm0.028.
\end{equation}
Taking into account the assumption that
$\Delta\alpha_{\rm{norm,PDF}}=0$, it follows immediately that the
two components of the effective multifractality of the original
volatility are
\begin{equation}
 \Delta\alpha_{\rm{PDF}} = \Delta\alpha_{\rm{eff}}-\Delta\alpha_{\rm{norm,eff}}=0.081\pm0.033
 \label{Eq:Original:dA:PDF}
\end{equation}
and
\begin{equation}
 \Delta\alpha_{\rm{NL}} = \Delta\alpha_{\rm{norm,NL}}=\Delta\alpha_{\rm{norm,eff}}=0.067\pm0.028.
 \label{Eq:Original:dA:NL}
\end{equation}
We note that, in the above framework, the nonlinearity component
does not change when the PDF is replaced.

\section{Impact of PDF}
\label{S1:ImpactPDF}

In order to further confirm that the PDF of the time series has
crucial influence on the multifractality, we investigate two
families of distributions with fat tails. The first one is a family
of Student's t distributions
\begin{equation}
 p(r_{\rm{t}}) = \frac{\Gamma\left(\frac{\gamma+1}{2}\right)}{\sqrt{\gamma\pi}\Gamma(\frac{\gamma}{2})}
        \left[1+\frac{(r_{\rm{t}}-\mu)^2}{\gamma}\right]^{-(\gamma+1)/2},
 \label{Eq:pdf:Student}
\end{equation}
which have power-law tails with exponent $\gamma$. In our
simulation, $\gamma$ ranges from 3 to 10 with a spacing step of 0.5.
For each tail exponent $\gamma$, we generate three types of
surrogates for the volatility $v(t)$ denoted as
$v_{\rm{t}}(t)=|r_{\rm{t}}(t)|$ which has the same temporal
correlations as $v(t)$, $v_{\rm{t,LM}}(t)$ which has the same linear
correlations as $v(t)$ while any nonlinear correlations are
eliminated, and $v_{\rm{t,SF}}(t)$ which is the shuffled data of
$v_{\rm{t}}(t)$. We have generated 100 series of $v_{\rm{t}}(t)$
which has the same length as $v(t)$. For each $v_{\rm{t}}(t)$, we
generate two surrogates $v_{\rm{t,LM}}(t)$ and $v_{\rm{t,LM}}(t)$.
The multifractal spectrum of each time series is determined.

\begin{figure}[htb]
 \centering
 \includegraphics[width=8cm]{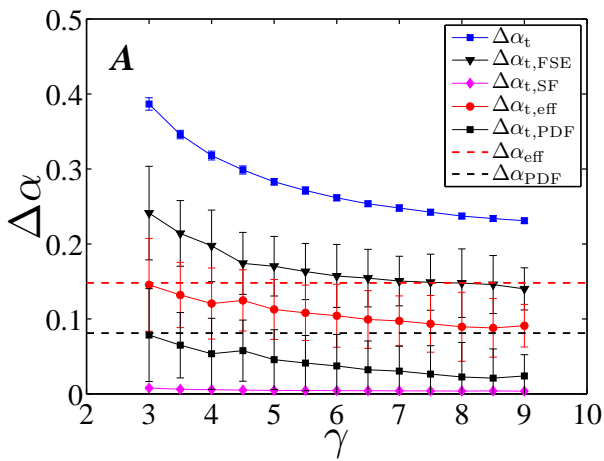}
 \caption{\label{Fig:MFComp:Volatility:PDF:t} (Color online) Dependence of the singularity width $\Delta\alpha$ as a function of the exponent $\gamma$ for surrogate time series with Student's $t$ distribution. The error bars are the standard deviations for the 100 surrogate series.}
\end{figure}

The apparent singularity width $\Delta\alpha_{\rm{t}}$, the
finite-size effect component $\Delta\alpha_{\rm{t,FSE}}$, the
effective singularity width $\Delta\alpha_{\rm{t,eff}}$, the PDF
component $\Delta\alpha_{\rm{t,PDF}}$ are calculated and illustrated
in Fig.~\ref{Fig:MFComp:Volatility:PDF:t} as a function of
the tail exponent $\gamma$. All the quantities decreased with
$\gamma$, implying that the multifractality is stronger when the
distribution is fatter (characterized by smaller $\gamma$). We also
plot $\Delta\alpha_{\rm{eff}}$ and $\Delta\alpha_{\rm{PDF}}$ for the
original volatility as horizontal lines, both of which intersect
with the $\Delta\alpha_{\rm{t,eff}}(\gamma)$ and
$\Delta\alpha_{\rm{t,PDF}}(\gamma)$ curves at $\gamma=3$. For
$\gamma=3$, we have $\Delta\alpha_{\rm{t}}=0.387\pm0.008$ for
$v_{\rm{t}}(t)$, $\Delta\alpha_{\rm{t,FSE}}=0.241\pm0.062$ for
$v_{\rm{t,LM}}(t)$, $\Delta\alpha_{\rm{t,SF}}=0.008\pm0.002$ for
$v_{\rm{t,SF}}(t)$, $\Delta\alpha_{\rm{t,eff}}=0.145\pm0.062$ and
$\Delta\alpha_{\rm{t,PDF}}=0.078\pm0.062$. This striking feature is
consistent with the inverse cubic law of financial returns
\cite{Gopikrishnan-Meyer-Amaral-Stanley-1998-EPJB}, which also holds
for daily DJIA returns \cite{Malevergne-Sornette-2006}.

The second one is a family of Weibull distributions
\begin{equation}
 p(v_{\rm{wbl}}) = \beta v_{\rm{wbl}}^{\beta-1}e^{-v_{\rm{wbl}}^\beta},
 \label{Eq:pdf:Weibull}
\end{equation}
where the shape parameter $\beta$ describes the heaviness of the tails and we require that $\beta\leqslant1$. In our simulation, $\beta$ varies from 0.4 to 1.0 with a spacing step of 0.1. For each $\beta$, 100 time series of the same length of $v(t)$ are generated, which are manipulated to generate $v_{\rm{wbl}}$ with the same rank ordering of $v(t)$ and $v_{\rm{wbl,LM}}$ only with linear correlations.

\begin{figure}[t]
 \centering
 \includegraphics[width=8cm]{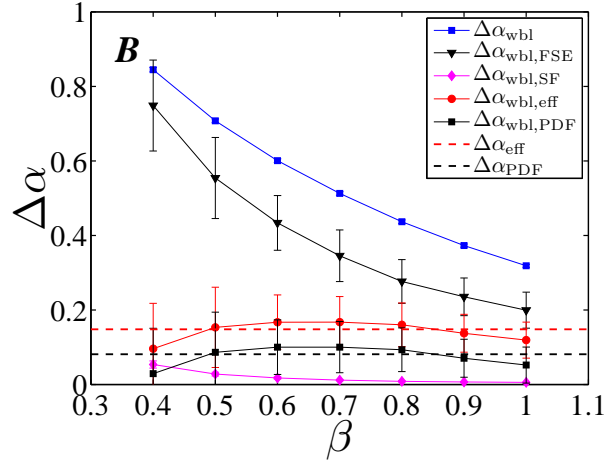}
 \caption{\label{Fig:MFComp:Volatility:PDF:wbl} (Color online) Dependence of the singularity width $\Delta\alpha$ as a function of the exponent $\beta$ for surrogate time series with Weibull distribution. The error bars are the standard deviations for the 100 surrogate series.}
\end{figure}

The results are shown in Fig.~\ref{Fig:MFComp:Volatility:PDF:wbl}. We find that $\Delta\alpha_{\rm{wbl}}$, $\Delta\alpha_{\rm{wbl,FSE}}$ and $\Delta\alpha_{\rm{wbl,SF}}$ decrease with $\beta$, while $\Delta\alpha_{\rm{wbl,eff}}$ and $\Delta\alpha_{\rm{wbl,PDF}}$ are concave functions with respect to $\beta$. For small $\beta$ values, $\Delta\alpha_{\rm{wbl,eff}}$ and $\Delta\alpha_{\rm{wbl,PDF}}$ increase with $\beta$, which may be caused by the fact that there is a lack of statistics so that the fluctuations of $\Delta\alpha_{\rm{wbl,FSE}}$ are markedly large. Indeed, the error bar decreases with increasing $\beta$. It is worth noting that the fluctuations of $\Delta\alpha_{\rm{t}}$ and $\Delta\alpha_{\rm{wbl}}$ are much less than the corresponding $\Delta\alpha_{\rm{t,FSE}}$ and $\Delta\alpha_{\rm{wbl,FSE}}$. The large error bars of the PDF components and the nonlinearity components are caused by the large error bars of the corresponding finite-size effect components. In addition, in the case of exponential distribution with $\beta=1$, we have
$\Delta\alpha_{\rm{exp}}=0.318\pm0.004$ for $v_{\rm{exp}}(t)$, $\Delta\alpha_{\rm{exp,FSE}}=0.198\pm0.005$ for $v_{\rm{exp,LM}}(t)$, and  $\Delta\alpha_{\rm{exp,SF}}=0.006\pm0.001$ for $v_{\rm{exp,SF}}(t)$. It follows that $\Delta\alpha_{\rm{exp,eff}}=0.120\pm0.005$ and $\Delta\alpha_{\rm{exp,PDF}}=0.053\pm0.005$.

\section{Summary}
\label{S1:Summary}

We have proposed a method to decompose the apparent multifractality into three components associated with the finite-size effect, the nonlinearity, and the fat-tailed PDF. The finite-size effect component can be determined by the singularity width of surrogate data with the same PDF and linear correlations as the original time series after eliminating its nonlinear correlations. The PDF component can be calculated as the difference between the effective multifractality of the original series and the surrogate data drawn from a normal distribution with the linear and nonlinear correlations preserved. The impact of the fat-tailed PDF on the effective multifractality is the outcome of the coupling of the PDF and the nonlinearity since the shuffled data are monofractal with vanishing singularity width. The method proposed in this paper can be applied to investigate the presence of multifractality in time series and, if any, determine its components.

It is necessary to clarify the marked differences between the current work and Ref.~\cite{Zhou-2009-EPL}. First, the current work investigates financial volatility, while Ref.~\cite{Zhou-2009-EPL} investigates financial returns, although the same data set of the daily DJIA index is used in both papers. Second, the partition function approach is used in this work, while the multifractal detrended fluctuation analysis is adopted in Ref.~\cite{Zhou-2009-EPL}. Third, the results are quite different. Ref.~\cite{Zhou-2009-EPL} shows that the multifractal spectrum width  $\Delta\alpha_{\rm{SF}}$ of the shuffled data is comparable to the apparent singularity width $\Delta\alpha$, which means that the long-term correlations have minor impact on $\Delta\alpha$ and the fat-tailed PDF plays a major role. In contrast, in Figs.~\ref{Fig:MFComp:Volatility:Correlation}, \ref{Fig:MFComp:Volatility:PDF:norm}, \ref{Fig:MFComp:Volatility:PDF:t} and \ref{Fig:MFComp:Volatility:PDF:wbl} of the current paper, we have shown that
\begin{equation}
 \Delta\alpha_{\rm{SF}} \approx 0
 \label{Eq:SI:dA:SF}
\end{equation}
for shuffled time series with different PDFs. It is clear that
\begin{equation}
 \Delta\alpha_{\rm{PDF}} \neq
 \Delta\alpha_{\rm{SF}}-\Delta\alpha_{\rm{norm,SF}},
 \label{Eq:SI:dA:SF:PDF}
\end{equation}
where $\Delta\alpha_{\rm{SF}}$ is the singularity width of the shuffled data $v_{\rm{SF}}$ and $\Delta\alpha_{\rm{norm,SF}}$ is the shuffled data $v_{\rm{norm,SF}}$. This suggests that the contribution of the PDF is coupled with the nonlinearity. If there is no nonlinearity, the fat-tailedness of the PDF will not introduce any multifractality. In addition, with the presence of nonlinearity, the PDF does have impact on the effective multifractality.

As a last note, it is well-established that the phase-randomized surrogates of heartbeat series from healthy subjects have a narrower width of singularity ($\Delta\alpha_{\rm{norm,LM}}\approx0.1$) compared to its apparent multifractal width ($\Delta\alpha\approx0.45$) based on the wavelet transform method \cite{Ivanov-Amaral-Goldberger-Havlin-Rosenblum-Struzik-Stanley-1999-Nature,Ivanov-Amaral-Goldberger-Havlin-Rosenblum-Stanley-Struzik-2001-Chaos}, which is in excellent agreement with our results.

\begin{acknowledgments}
We are grateful to Stanislaw Dro\.zd\.z for invaluable discussions. This work was partially supported by the ``Shu Guang'' project supported by Shanghai Municipal Education Commission and Shanghai Education Development Foundation under grant 2008SG29 and the Program for New Century Excellent Talents in University sponsored the Ministry of Education of People's Republic of China under grant NCET-07-0288.
\end{acknowledgments}

\bibliography{E:/Papers/Auxiliary/Bibliography} 

\end{document}